\documentclass[
    ,final            % use final for the camera ready runs
  ]
  {aipproc}

\layoutstyle{6x9}

\setcitestyle{numbers}
\begin{document}

\title{Testing for three-body quark forces in $L=1$ excited baryons }


\author{Dan Pirjol}{
  address={ National
Institute for Physics and Nuclear Engineering, Department of Particle
Physics, \\  077125 Bucharest, Romania}
}

\author{Carlos Schat\thanks{Speaker, {\it XI Hadron Physics}, March 21-26, 2010, S\~ao Paulo, Brazil} }
{
  address={Department of Physics and Astronomy, Ohio University, Athens, Ohio 45701, USA \\
           Departamento de F\'{\i}sica, FCEyN,
Universidad de Buenos Aires, Ciudad Universitaria, Pab.1, (1428) Buenos
Aires, Argentina} 
}

\begin{abstract}
We discuss the matching of the quark model to the effective mass operator of 
the $1/N_c$ expansion using the permutation group $S_N$. 
As an illustration of the general procedure we perform the matching 
of the Isgur-Karl model
for the spectrum of the negative parity
$L=1$ excited baryons. Assuming the most general two-body quark Hamiltonian, we derive two correlations
among the masses and mixing angles of these states which should hold in any quark model.
These correlations constrain the mixing angles and can be
used to test for the presence of three-body quark forces. 
\end{abstract}

\maketitle

%%%%%%%%%%%%%%%%%%%%%%%%%%%%%%%%%%%%%%%%%%%%
%% MAINMATTER
%%%%%%%%%%%%%%%%%%%%%%%%%%%%%%%%%%%%%%%%%%%%

\section{Introduction}

Quark models provide a simple and  intuitive picture of the physics of 
ground state baryons and their excitations \cite{De Rujula:1975ge,Isgur:1977ef}. 
An alternative description is provided by the $1/N_c$ expansion,
which is a systematic approach to the study of 
baryon properties \cite{Dashen:1993jt}.
This program can be realized in terms of a quark operator expansion, 
which gives rise to a physical picture similar to the one of the 
phenomenological quark models, but is closer connected to QCD.  In this 
context quark models gain additional significance. 

The $1/N_c$ expansion has been applied both to the ground state and excited nucleons
\cite{Goity:1996hk,Pirjol:1997bp,Carlson:1998vx,Schat:2001xr,Goity:2003ab,Pirjol:2003ye,Matagne:2004pm}.
In the system of negative parity $L=1$ excited baryons this approach has 
yielded a number of interesting insights for the spin-flavor structure of the 
quark interaction.

In a recent paper \cite{Pirjol:2007ed} we showed how to match an 
arbitrary quark model Hamiltonian onto the operators of the $1/N_c$ expansion, 
thus making the connection between these two physical pictures.
This method makes use of the transformation of the states and operators
under $S_N^{\rm sp-fl}$, the permutation group of $N$ objects acting
on the spin-flavor degrees of the quarks. This is similar to the 
method discussed in Ref.~\cite{Collins:1998ny} for $N_c=3$ in terms of
$S_3^{\rm orb}$, the permutation group of three objects acting on the
orbital degrees of freedom.

The main result of \cite{Pirjol:2007ed} can be summarized as follows:
consider a two-body quark Hamiltonian $V_{qq} = \sum_{i<j} O_{ij} R_{ij}$,
where $O_{ij}$ acts on the spin-flavor quark degrees of freedom and $R_{ij}$
acts on the orbital degrees of freedom. Then the hadronic matrix elements of
the quark Hamiltonian on a baryon state $|B\rangle$ contains only the projections
$O_\alpha$ of $O_{ij}$ onto a few irreducible representations of $S_N^{\rm sp-fl}$ 
and can be factorized as $\langle B |V_{qq}|B\rangle = \sum_\alpha C_\alpha 
\langle O_\alpha\rangle$. The coefficients $C_\alpha$ are related to
reduced matrix elements of the orbital operators $R_{ij}$, and are
given by overlap integrals of the quark model wave functions.
The matching procedure has been discussed in detail 
for the Isgur-Karl (IK) model in Ref.~\cite{Galeta:2009pn} providing a simple example 
of this general formalism.
 
In Ref.~\cite{Pirjol:2008gd,PirSch:2010a} we used the general $S_N$ approach to study 
the predictions of the quark model with the most general two-body quark interactions, and to
obtain information about the spin-flavor structure of the quark interactions
from the observed spectrum of the $L=1$ negative parity baryons. This talk
summarizes the main ideas and emphasizes their relevance as a possible test for 
three-body forces in excited baryons. 

\section{The mass operator of the Isgur-Karl model}
\label{IKV}

The Isgur-Karl model is defined by the quark Hamiltonian
\begin{eqnarray}
{\cal H}_{IK} = {\cal H}_0 + {\cal H}_{\rm hyp}  \,, 
\end{eqnarray}
where
${\cal H}_0$ contains the confining potential and kinetic terms of the quark
fields, and is symmetric under spin and isospin. The hyperfine
interaction ${\cal H}_{\rm hyp}$ is given by 
\begin{eqnarray}\label{HIK} 
{\cal H}_{\rm hyp} = A \sum_{i<j}\Big[ \frac{8\pi}{3} \vec s_i \cdot \vec s_j
\delta^{(3)}(\vec r_{ij}) + \frac{1}{r_{ij}^3} (3\vec s_i \cdot \hat r_{ij} \
\vec s_j \cdot \hat r_{ij} - \vec s_i\cdot \vec s_j) \Big] \,, 
\end{eqnarray} 
where $A$ determines the strength of the interaction, and
$\vec r_{ij} = \vec r_i - \vec r_j$ is the distance between quarks
$i,j$.  The first term is a local spin-spin interaction, and the second
describes a tensor interaction between two dipoles. This interaction
Hamiltonian is an approximation to the gluon-exchange interaction,
neglecting the spin-orbit terms\footnote{In Ref.\cite{Isgur:1977ef} A is taken as $A=\frac{2 \alpha_S}{3 m^2}$.}. 

 In the original formulation of the 
IK model \cite{Isgur:1977ef} the confining forces are harmonic.  
We will derive in the following the form 
of the mass operator without making any assumption on the shape of the confining quark forces. 
We refer to this more general version of the model as IK-V(r).

The $L=1$ quark model states include the following SU(3) multiplets: 
two spin-1/2 octets $8_\frac12, 8'_\frac12$, two spin-3/2 octets $8_\frac32, 8'_\frac32$,
one spin-5/2 octet $8'_\frac52$, two decuplets $10_\frac12, 10_\frac32$ and two singlets
$1_\frac12, 1_\frac32$. States with the same quantum numbers mix. For the 
$J=1/2$ states we define
the relevant mixing angle $\theta_{N1}$ 
in the nonstrange sector as
\begin{eqnarray}
N(1535) & = &  \cos\theta_{N1} N_{1/2} + \sin\theta_{N1} N'_{1/2} \,, \\
N(1650) & = &  -\sin\theta_{N1} N_{1/2} + \cos\theta_{N1} N'_{1/2} 
\end{eqnarray}
and similar equations for the $J=3/2$ states, which define a second mixing angle 
$\theta_{N3}$.

We find \cite{Galeta:2009pn} that the most general
mass operator in the IK-V(r)  model depends only on three unknown orbital overlap
integrals, plus an additive constant $c_0$ related to the matrix element
of ${\cal H}_0$, and can be written as 
\begin{eqnarray}\label{IKMass} 
\hat M =
c_0 + a  S_c^2 + b L_2^{ab} \{ S_c^a\,, S_c^b\}  + c L_2^{ab} \{
s_1^a\,, S_c^b\} \,,
\end{eqnarray} 
where the spin-flavor operators are
understood to act on the state $|\Phi(SI)\rangle$ constructed as a
tensor product of the core of quarks 2,3 and the `excited' quark 1, as given in 
\cite{Carlson:1998vx,Pirjol:2007ed}.
  The
coefficients are given by
\begin{eqnarray}\label{coefa}
&& a = \frac12 \langle R_S\rangle \,,\, \  
b = \frac{1}{12} \langle Q_S\rangle - \frac16 \langle Q_{MS}\rangle \,,\,  \  
c = \frac16 \langle Q_S\rangle + \frac16 \langle Q_{MS}\rangle \,\,. \label{coefc}
\end{eqnarray}
The reduced matrix elements $R_S,Q_S,Q_{MS}$ for the orbital part of the interaction contain 
the unknown spatial dependence and are defined in Refs.~\cite{Carlson:1998vx,Galeta:2009pn}.
Evaluating the matrix elements using Tables~II,~III in
Ref.~\cite{Carlson:1998vx} we find the following explicit result for
the mass matrix 
\begin{eqnarray}
M_{1/2} &=& 
\left( 
\begin{array}{cc}
c_0 + a                   &  -\frac53 b + \frac{5}{6}c \\ 
-\frac53 b + \frac{5}{6}c & c_0 + 2a + \frac53(b+c)\\ 
\end{array} 
\right) \,, \\ 
M_{3/2} &=& 
\left(
\begin{array}
{cc} c_0 + a                                  &  \frac{\sqrt{10}}{6} b -\frac{\sqrt{10}}{12}c \\ 
\frac{\sqrt{10}}{6} b - \frac{\sqrt{10}}{12}c & c_0 + 2a - \frac43(b+c)\\ 
\end{array}
\right) \,, \\ 
M_{5/2}      &=& c_0 + 2a +\frac13 (b+c) \,, \\ 
\Delta_{1/2} &=& \Delta_{3/2} = c_0 + 2a \,.
\end{eqnarray} 
\begin{figure}[t!]
\includegraphics[width=8.0cm]{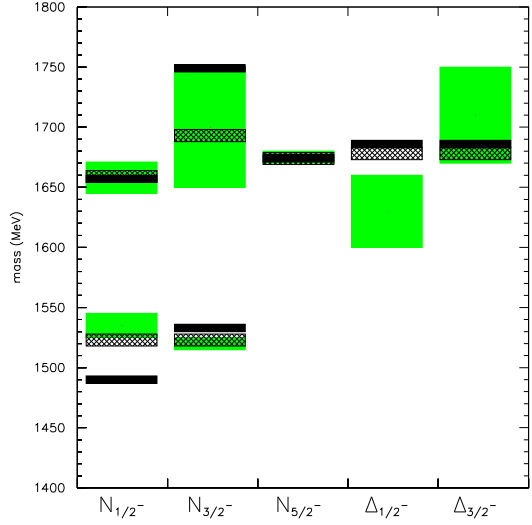}
\caption{Masses predicted by the IK model (black bars), by the IK-V(r) model (hatched bars) 
and the experimental masses (green boxes) from Ref.~\cite{Amsler:2008zzb}. }
\label{fig:masses}
\end{figure}
Computing the reduced matrix elements with the interaction given by Eq.~(\ref{HIK}),  one finds that the
reduced matrix elements  in the IK model with harmonic oscillator
wave functions are all related and can be expressed in terms of the single parameter $\delta$ as
\begin{eqnarray} 
\langle Q_{MS}\rangle = \langle Q_S\rangle = -
\frac35 \delta \qquad ; \qquad \langle R_S\rangle = \delta \,.
\end{eqnarray} 
This gives a relation among
the coefficients $a,b,c$ of the mass matrix Eq.~(\ref{IKMass}) 
\begin{eqnarray} \label{coefIK}
a = \frac12 \delta \,, \qquad b = \frac{1}{20} \delta \,, \qquad 
c = - \frac15 \delta \,.
\end{eqnarray}
We recover the well known result that in the harmonic oscillator model,  the entire spectroscopy of 
the $L=1$ baryons is fixed by one
single constant $\delta=M_\Delta-M_N \sim 300 \  {\rm MeV}$, along with an overall additive constant $c_0$, and
the model becomes very predictive.

In Fig.~\ref{fig:masses} we show the result of a best fit of $a,b,c$ in the IK-V(r) model together 
with the predictions of the IK model. The IK-V(r) spectrum is the best fit possible for a 
potential model with the spin-flavor interaction given in Eq.~(\ref{HIK}).

\section{The most general two-body quark Hamiltonian}
\label{Sec:Hamiltonian}

The most general two-body quark interaction Hamiltonian in the constituent
quark model can be written in generic form as $V_{qq} = \sum_{i<j} V_{qq}(ij)$
with
\begin{eqnarray}
\label{2}
V_{qq}(ij) &=& \sum_k f_{0,k}(\vec r_{ij}) O_{S,k}(ij) + 
f_{1,k}^a(\vec r_{ij}) O_{V,k}^a(ij)  + 
f_{2,k}^{ab}(\vec r_{ij}) O_{T,k}^{ab}(ij)\,, 
\end{eqnarray}
where $a,b=1,2,3$ denote spatial indices, 
$O_{S}, O_V^a, O_T^{ab}$ act on spin-flavor, and $f_k(\vec r_{ij})$ are 
orbital functions. Their detailed form is unimportant
for our considerations. The scalar, spin-orbit and tensor parts of the interaction yield factors of
$\mathbf{1}, L^i, L_2^{ij} = \frac12 \{L^i, L^j\} - \frac13\delta^{ij} L(L+1)$, which are
coupled to the spin-flavor part of the interaction as shown in Table~I  of Ref.~\cite{Pirjol:2008gd}

Following Refs.~\cite{Pirjol:2007ed,Pirjol:2008gd}  one finds that the most general form of the effective
mass operator in the presence of  these two-body quark interactions is a linear 
combination of 10 nontrivial spin-flavor operators
\begin{eqnarray}\label{10Ops}
& & O_1 = T^2\,,\,\, O_2 = \vec S_c^2\,,\,\, O_3 = \vec s_1\cdot \vec S_c\,,
\,\, O_4 = \vec L\cdot \vec S_c\,,\,\, O_5 = \vec L\cdot \vec s_1\,,\,\,
O_6 = L^i t_1^a G_c^{ia}\,,\nonumber \\
& & O_7 = L^i g_1^{ia} T_c^a\,,\,\,
O_8 = L_2^{ij} \{ S_c^i, S_c^j\} \,,\,\,
O_9 = L_2^{ij} s_1^i S_c^j \,,\,\, O_{10} = L_2^{ij} g_1^{ia} G_c^{ja} \,
\end{eqnarray}
and the unit operator.

It turns out that the 11 coefficients $C_{0-10}$ contribute to the mass
operator of the negative parity $N^*$ states only in 9 independent
combinations: $C_0,C_1 - C_3/2, C_2+C_3,
C_4, C_5, C_6, C_7, C_8+C_{10}/4, C_9 - 2C_{10}/3$.
This implies the existence of two universal 
relations among the masses of the 9 multiplets plus the two mixing angles, which must
hold in any quark model containing only two-body quark interactions.

\begin{figure}[t!]
\includegraphics[width=8.0cm]{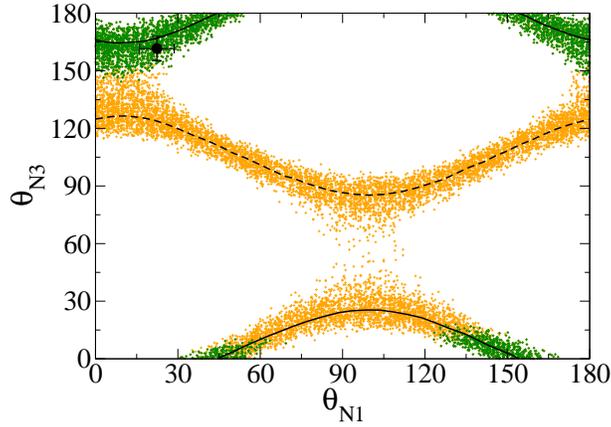}
\caption{Correlation in the $(\theta_{N1}, \theta_{N3})$ plane from the
quark model with the most general two-body quark interactions. }
\label{fig:corr}
\end{figure}

The first universal relation involves only the nonstrange hadrons, and requires only 
isospin symmetry. It can be expressed as a correlation among the two mixing
angles $\theta_{N1}$ and $\theta_{N3}$ (see Fig.~\ref{fig:corr} )
\begin{eqnarray}\label{correl}
&& \frac{1}{2} (N(1535) + N(1650)) + \frac{1}{2}(N(1535)-N(1650))
(3 \cos 2\theta_{N1} + \sin 2\theta_{N1}) \\
&& - \frac{7}{5} (N(1520) + N(1700)) + (N(1520) - N(1700))
\Big[ - \frac{3}{5} \cos 2\theta_{N3} + \sqrt{\frac52} \sin 2\theta_{N3}\Big] \nonumber \\
&& = -2 \Delta_{1/2} + 2 \Delta_{3/2} - \frac{9}{5} N_{5/2} \nonumber\,. 
\end{eqnarray}
This correlation holds also model independently in the $1/N_c$ expansion,
up to corrections of order $1/N_c^2$, since for non-strange states the 
mass operator to order $O(1/N_c)$ \cite{Carlson:1998vx,Schat:2001xr} is generated by
 the operators in Eq.~(\ref{10Ops}).
An example of an operator which violates this correlation is $L^i g^{ja} \{ S_c^j\,,
G_c^{ia}\}$, which can be introduced by three-body quark forces. 
In Fig.\ref{fig:corr} we show the two solutions correlating the two mixing angles, 
where the solid line and the dashed line are obtained replacing the  
central values of the baryon masses in Eq.~(\ref{correl}). 
These lines are expanded into bands given by the  scatter plot when the
experimental errors in the masses are taken into account. 
The second universal relation expresses the spin-weighted SU(3) singlet mass
$\bar \Lambda = \frac16  (2\Lambda_{1/2} + 4 \Lambda_{3/2})$ 
in terms of the nonstrange hadronic parameters and can be found in Ref.~\cite{Pirjol:2008gd}. 
The green area in 
 Fig.~\ref{fig:corr} shows the allowed region for $(\theta_{N1},
\theta_{N3})$ compatible with both relations and singles out the solid line as the 
preferred solution.

On the same plot we show also, as a black dot with error bars, 
the values of the mixing angles obtained in Ref.~\cite{Goity:2004ss}
from an analysis of  the $N^*\to N\pi$ strong decays and in  Ref.~\cite{Scoccola:2007sn}
from photoproduction amplitudes. 

These angles are  in good agreement with the  
correlation Eq.~(\ref{correl}), and provide no evidence for the presence of spin-flavor dependent three-body quark 
interactions, within errors. It would be interesting to narrow down the errors on masses 
and mixing angles, and also compare with the upcoming results of lattice calculations for 
these excited states, to see if violations of the correlation given by Eq.~(\ref{correl}) 
become apparent. 

\section{Conclusions}

We presented the Isgur-Karl model mass operator in a form that makes the connection with 
the $1/N_c$ operator expansion clear. This simple and explicit calculation (for details 
see Ref.~\cite{Galeta:2009pn}) should 
serve as an illustration of the general matching procedure discussed in Ref.~\cite{Pirjol:2007ed}
using the permutation group.

We used the more general matching procedure
\cite{Pirjol:2008gd}
 to saturate the contribution of all 
possible two-body forces
to the masses of the negative parity $L=1$ excited baryons, without 
making any assumptions about the orbital hadronic wave functions. We derived two
universal correlations among masses and mixing angles, which 
will be broken by the presence of three-body forces, and could 
be used to set bounds on their strength given a more precise determination 
of all the masses and mixing angles for the negative parity $L=1$ baryons.

\begin{theacknowledgments}

The work of C.S. was supported by CONICET and partially supported by  the U.~S.  Department of Energy, Office of
Nuclear Physics under contract No. DE-FG02-93ER40756
with Ohio University.

\end{theacknowledgments}

\end{document}